\documentclass[conference,10pt]{IEEEtran}
\IEEEoverridecommandlockouts

\usepackage{epsfig,rotating,setspace,latexsym,amsmath,epsf,amssymb,amsfonts,bm,theorem,cite,algorithm,graphicx,epsf,authblk,epstopdf,color,algpseudocode,bbm}

\usepackage{pgfplots}
\usepackage{filecontents}

\usepackage{texdef2020local}

\usepackage[algo2e]{algorithm2e}
\usepackage{tikz}
\usetikzlibrary{automata,arrows,positioning,calc,shapes.multipart,chains,shapes,decorations}

\newtheorem{theorem}{Theorem}

\newcommand{\AoI}{\overline{\texttt{AoI}}}
\newcommand{\VoI}{\overline{\texttt{VoI}}}

\newcommand{\AGE}{A}
\newcommand{\VALUE}{V} 
\newcommand{\TIME}{T}
\newcommand{\fyi}[1][i]{\hat{f}_{#1}}
\newcommand{\betabar}{\bar{\beta}}
\newcommand{\thresh}[1][\theta]{\tau(#1)}
\newcommand{\ythresh}[1][i]{\bar{y}_{#1}(\theta)}

\allowdisplaybreaks

\begin{document}

\title{Age and Value of Information Optimization for Systems with Multi-Class Updates\thanks{This work was supported by the U.S. National Science Foundation under Grants CNS 21-14537 and ECCS 21-46099.}
}

\author[1]{Ahmed Arafa}
\author[2]{Roy D. Yates}
\affil[1]{\normalsize Department of Electrical and Computer Engineering, University of North Carolina at Charlotte}
\affil[2]{\normalsize Department of Electrical and Computer Engineering, Rutgers University}

\maketitle

\begin{abstract}
Received samples of a stochastic process are processed by a server for delivery as updates to a  monitor. Each sample belongs to a class that specifies a distribution for its processing time and a function that describes how the value of the processed update decays with age at the monitor. The class of a sample is identified when the processed update is delivered. The server implements a form of M/G/1/1 blocking queue; samples arriving at a busy server are discarded and samples arriving at an idle server are subject to an admission policy that depends on the age and class of the prior delivered update. For the delivered updates, we characterize the average age of information (AoI) and average value of information (VoI). We derive the optimal stationary policy that minimizes  the convex combination  of the AoI and (negative) VoI. It is shown that the policy has a threshold structure, in which a new sample is allowed to arrive to the server only if the previous update's age and value difference surpasses a certain threshold that depends on the specifics of the value function and system statistics.
\end{abstract}

\section{Introduction}

Consider a system in which time-stamped raw data samples are processed into updates for a monitor. The server and the monitor are co-located, and each sample incurs a {\em processing time} before being delivered as an update to the monitor. Some updates can be processed quickly, while others require longer service times. Updates with long processing times may be unusually important or valuable at the monitor. Occasionally, an update may have a very long service time or be of exceptional importance.  

One such example is an augmented reality (AR) system in which images are processed and analyzed, and an update in the form of an image augmentation is delivered to the monitor.  In the AR system, time-stamped images are samples that arrive at the input  of a processing system. When an input job is processed, the output, namely an image augmentation, represents an update. The time-stamp of the update is the time-stamp of the image from which it was derived. 

For a second example, consider a camera system monitoring  an urban crosswalk. The video frame updates require processing to generate bounding boxes that correspond to pedestrians in the crosswalk.  As the number of pedestrians increases, the video frames require more processing time to identify bounding boxes. However, in the context of pedestrian safety, images with more pedestrians are more important and have greater value.  

Object recognition is typically  a key step  in these image processing applications. To find a particular object in a given image input, the system extracts key feature points from the image input, and then matches all the feature points with those of the particular object. With a high matching ratio, it assumes the object has been detected \cite{Zhang-Xu-ipdps2008}. However, when there is a large number of objects in the input,  there will be numerous feature points and this will increase the matching complexity and thus the processing time. The value of an image is likely to be increasing with its number of classified objects.

To model such situations, samples are categorized into $M$ different classes.  A sample belongs to class $i$ with probability $p_i$.  Each class $i$ sample has a class-dependent processing time and  carries a {\it value} $\nu_i(t)$ at time $t$. Such value is only revealed at the monitor after being processed by the server. 

Both timeliness and value of the updates are important. We evaluate the performance of the updating system using both age of information (AoI) \cite{yates_age_1} and value of information (VoI) \cite{ephremides_age_non_linear} metrics. 
In particular, we develop a class of server policies that minimize a convex combination of AoI and (negative) VoI. This enables us to compare the characteristics of AoI and VoI minimization as well as tradeoffs between AoI and VoI.

In prior work on updates with non-memoryless service times \cite{bedewy-aoi-multihop, inoue-aoi-general-formula-fcfs, najm-age-mg11-harq}, it has been observed that average AoI  can be reduced substantially by a simple preemption-in-service mechanism.  Preemption can replace an update that becomes stale while in service with a fresh update and this can substantially reduce the AoI. However, in the context of multiple classes of updates, it is unclear whether this is a desirable approach. Specifically, if updates with long processing times are the important high-value updates, preemption will result in low VoI at the monitor. Moreover, VoI alters how we should process updates.  Specifically, the system is earning a reward over time for update $k-1$ following its delivery. In that same time period, the system is either processing update $k$ or waiting to begin the processing of update $k$. The key question is 
\begin{center}
{\it how long should the system accrue value from update $k-1$ before initiating the process to deliver update $k$?}
\end{center}

Based on these considerations, we believe that reducing AoI via preemption in service may be inappropriate for some applications.
In particular, preemption in service will be  biased in that updates with longer service times are  more likely to be preempted. On the other hand, queueing of updates also remains undesirable; timeliness is improved when the system avoids processing updates that have become stale in a queue. Hence, this work focuses on an M/G/1/1 queueing model  with blocking: if the server is busy, new arrivals are blocked and discarded. This mechanism avoids queueing but also avoids a bias against jobs with long service times. Whether an arriving job goes into service (or is blocked) is independent of its service class. Moreover, once a job goes into service, it is guaranteed to finish processing, independent of its class.

\subsection{Related Work}

Prior work \cite{Huang-Modiano-isit2015} on the AoI analysis of multi-class queueing systems has  examined peak AoI (PAoI) in  multiclass M/G/1 and M/G/1/1 queues. Each traffic class is described by its arrival rate, and the first and second moments of its service time, and arrival rates are optimized to minimize $\max_i C_i(A_i)$, where $C_i(A_i)$ is the cost of stream $i$ having PAoI of $A_i$. In a study of the average AoI for multiple streams arriving at an M/G/1/1 queue with preemption \cite{najm-age-mg11-preempt}, with all streams having the same general service time, it is shown that increasing the arrival rate for one stream/class can reduce its AoI, but at the  expense of increased AoI for other classes.

This work differs from these prior  M/G/1/1 studies in that updates belong to different service classes but they all originate from the same source. The overall update rate $\lambda$ is a controllable input but the probability $p_i$ that an arriving job is class $i$ is a property of the application scenario.  In the AR example, the service rates of the classes would depend on the complexity and variety of the scenario-specific images.

\subsection{Notation}

We define $[x]^+=\max(x,0)$. A random variable $R$ has probability density function $\pdf{R}{r}$, expected value $\E{R}$ and moment generating function (MGF) $\Phi_R(s)=\E{e^{sR}}$.

\section{System Model}

At the processor, a received  sample belongs to class $i$ with probability $p_i$. A class $i$ sample consumes time $\Yhat_i$ in processing.  Let $Y_k$ denote the processing time of the $k$th sample. For instance, if the $\Yhat_i$'s are exponential $(\lambda_i)$, then $Y_k$ would be hyperexponential. We assume that processing times and class identities are independent and identically distributed (i.i.d.) across samples.

Let $S_k$ and $D_k$ denote the sampling and delivery time of the $k$th sample/update, respectively, and let $i_k$ indicate its class, i.e., $i_k=i$ if sample $k$ belongs to class $i$. Hence,
\begin{align}
D_k=S_k+Y_k.
\end{align}

At the monitor, two quantities are observed over time: 1) the age-of-information (AoI),
\begin{align}
\Delta(t)\triangleq t-S_k,\quad D_k\leq t<D_{k+1};
\end{align}
and 2) the value-of-information (VoI), 
\begin{align}
V(t)=\nu_{i_k}(t-S_k),\quad D_k\leq t<D_{k+1}.
\end{align} 
That is, $\nu_{i_k}(\tau)$ denotes the value carried by the $k$th update when that update has age $\tau$. Note that $\nu_{i_k}(\tau)=\nu_i(\tau)$ in case update $k$ belongs to class $i$.

While our formulation allows for general forms of value functions $\nu_{i}(\tau)$, we will focus on {\em exponentially-decaying} value functions given by
\begin{align} \label{eq_exp-val}
\nu_{i}(\tau)=\nu_{i}e^{-\alpha_{i}\tau},\quad \tau\ge 0,
\end{align}
for some 
$\alpha_i\ge 0$. 
Thus the class $i$ of an update specifies its initial value $\nu_i$ (which we will often refer to as the value of a class $i$ update) and its value decay rate $\alpha_i$.  When $\alpha_i=0$, we obtain the special case of {\it fixed} value functions.

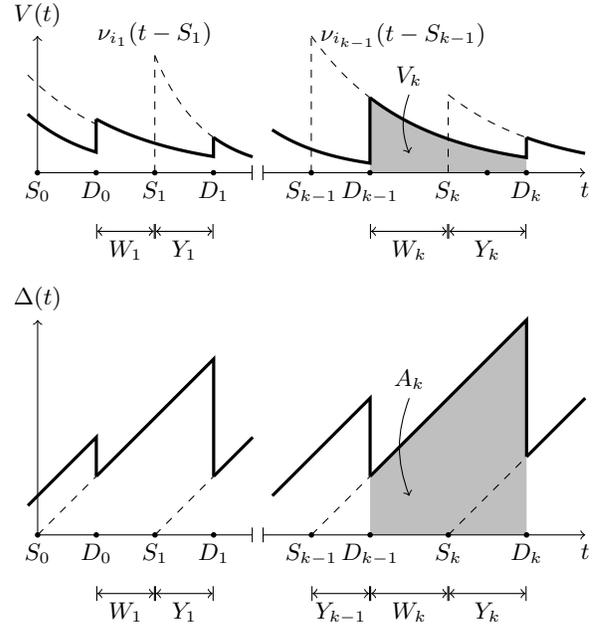
\begin{figure}
\begin{tabular}{c}
\begin{tikzpicture}[scale=0.26]
\pgfdeclaredecoration{arrows}{draw}{
\state{draw}[width=\pgfdecoratedinputsegmentlength]{%
  \path [every arrow subpath/.try] \pgfextra{%
    \pgfpathmoveto{\pgfpointdecoratedinputsegmentfirst}%
    \pgfpathlineto{\pgfpointdecoratedinputsegmentlast}%
   };
}}
\tikzset{every arrow subpath/.style={|<->|, draw}}
\fill [fill=lightgray,domain=17:25] plot (\x, {7*exp(-0.2*(\x-14))}) to (25,0) to (17,0) to (17,{7*exp(-0.2*(17-14) )});
\draw [<-|] (0,7) node [above] {\small $V(t)$} -- (0,0) -- (11,0);
\draw [|->] (11.5,0) -- (28,0) node [below] {\small$t$};
\fill
(3,0)  circle[radius=4pt](6,0)  circle[radius=4pt]
(9,0)  circle[radius=4pt]
(17,0)  circle[radius=4pt]
(14,0)  circle[radius=4pt]
(23,0)  circle[radius=4pt]
(25,0)  circle[radius=4pt];
\fill
(0,0) circle[radius=4pt] node [below] {\small$S_0$}
(3,0) node [below] {\small$D_0$}
(6,0) node [below] {\small$S_1$}
(9,0) node [below] {\small$D_1$}
(14,0) node [below] {\small$S_{k-1}$}
(17,0) node [below] {\small$D_{k-1}$}
(21,0) node [below] {\small$S_{k}$}
(25,0) node [below] {\small$D_{k}$};

\draw[ very thick, domain=-0.5:3] plot (\x, {3*exp(-0.3*(\x+0.5))}) -- (3, {5*exp(-0.2*3});
\draw[ thin, dashed, domain=-0.5:3] plot (\x, {5*exp(-0.2*(\x+0.5))});
\draw[ very thick, domain=3:9] plot (\x, {5*exp(-0.2*(\x))}) -- (9, {6*exp(-0.4*(9-6))});
\draw[thin, dashed] (6,0)--(6,6) node [above] {\small$\nu_{i_1}(t-S_1)$};
\draw[thin, dashed, domain=6:9] plot (\x, {6*exp(-0.4*(\x-6))});
\draw[ very thick, domain=9:11] plot (\x, {6*exp(-0.4*(\x-6))});

\draw[very thick, domain=12:17] plot (\x, {4*exp(-0.3*(\x-10))}) -- (17,{7*exp(-0.2*(17-14))});
\draw[ thin, dashed] (14,0)--(14,7) node [above,right] {\small$\nu_{i_{k-1}}(t-S_{k-1})$};
\draw[ thin, dashed, domain=14:17] plot (\x, {7*exp(-0.2*(\x-14))});

\draw[ very thick, domain=17:25] plot (\x, {7*exp(-0.2*(\x-14))}) -- (25,{4*exp(-0.2*(25-21))});
\draw[<-] (19,1) to [out=110,in=250] (19,4) node [above] {\small$V_k$};
\draw[ thin, dashed] (21,0)--(21,4);
\draw[ thin, dashed, domain=21:25] plot (\x, {4*exp(-0.2*(\x-21))});
\draw[ very thick, domain=25:28] plot (\x, {4*exp(-0.2*(\x-21))});
\path [decoration=arrows, decorate] (3,-3) to node [below] {\small$W_1$} ++(3,0) to node [below] {\small$Y_1$}  ++(3,0);
\path [decoration=arrows, decorate] (17,-3) to node [below] {\small$W_k$} ++(4,0) to node [below] {\small$Y_k$}  ++(4,0);
\end{tikzpicture}\\
\begin{tikzpicture}
[scale=0.26]
\pgfdeclaredecoration{arrows}{draw}{
\state{draw}[width=\pgfdecoratedinputsegmentlength]{%
  \path [every arrow subpath/.try] \pgfextra{%
    \pgfpathmoveto{\pgfpointdecoratedinputsegmentfirst}%
    \pgfpathlineto{\pgfpointdecoratedinputsegmentlast}%
   };
}}
\tikzset{every arrow subpath/.style={|<->|, draw}}
\fill [fill=lightgray] (17,0) to ++(0,3) to ++(8,8) to ++(0,-11);
\draw [<-|] (0,11) node [above] {\small $\Delta(t)$} -- (0,0) -- (11,0);
\draw [|->] (11.5,0) -- (28,0) node [below] {\small$t$};
\fill
(0,0) circle[radius=4pt]
(3,0)  circle[radius=4pt](6,0)  circle[radius=4pt]
(9,0)  circle[radius=4pt]
(17,0)  circle[radius=4pt]
(14,0)  circle[radius=4pt]
(21,0)  circle[radius=4pt]
(25,0)  circle[radius=4pt];
\draw
(0,0) node [below] {\small$S_0$}
(3,0) node [below] {\small$D_0$}
(6,0) node [below] {\small$S_1$}
(9,0) node [below] {\small$D_1$}
(14,0) node [below] {\small$S_{k-1}$}
(17,0) node [below] {\small$D_{k-1}$}
(21,0) node [below] {\small$S_{k}$}
(25,0) node [below] {\small$D_{k}$};

\draw[ very thick, domain=-0.5:3] plot (\x, {\x+2}) -- (3,3);
\draw[ thin, dashed, domain=0:3] plot (\x,\x);
\draw[very thick, domain=3:9] plot (\x,\x) -- (9,3);
\draw[thin, dashed] (6,0)--(9,3);
\draw[ very thick, domain=9:11] plot (\x,\x-6);
\draw[very thick, domain=12:17] plot (\x,\x-10) -- (17,3);
\draw[ thin, dashed] (14,0)--(17,3);
\draw[ very thick, domain=17:25] plot (\x,{\x-14}) -- (25,4);
\draw[<-] (19,2) to [out=110,in=250] (19,7) node [above] {\small$A_k$};
\draw[ thin, dashed, domain=21:25] plot (\x,\x-21);
\draw[ very thick, domain=25:28] plot (\x,\x-21);
\path [decoration=arrows, decorate] (3,-3) to node [below] {\small$W_1$} ++(3,0) to node [below] {\small$Y_1$}  ++(3,0);
\path [decoration=arrows, decorate] (14,-3) to node [below] {\small$Y_{k-1}$} ++(3,0) to node [below] {\small$W_k$} ++(4,0) to node [below] {\small$Y_k$}  ++(4,0);
\end{tikzpicture}
\end{tabular}
\caption{Sample paths of age $\Delta(t)$ and exponentially decaying value $V(t)$.}
\label{fig:samplepath}
\vspace{-.2in}
\end{figure}

An example sample path of AoI and VoI versus time is shown in Figure~\ref{fig:samplepath}.
 
Following the processing and delivery of update $k-1$, let $W_k$ denote the waiting time before the server starts processing the $k$th sample. We consider an {\it external arrivals} model, in which samples arrive as an exogenous rate $\lambda$ Poisson process. In this model, our choice is to either block/discard  an arriving sample or to admit and process that sample. Consequently, the waiting time $W_k$ between the delivery of update $k-1$ and the start of processing of sample $k$ is \begin{align}
\eqnlabel{exogenousW}
W_k=W'_k+X_k,    
\end{align} 
where $W'_k $ is a controlled waiting time during which arriving samples are blocked and $X_k$ is the uncontrolled  random exponential $(\lambda)$ time until a new exogenous sample arrives following the end of the controlled waiting/blocking period. After a sample gets admitted, any arriving sample while the server is busy processing will be blocked/discarded. 

In the limit as $\lambda\to\infty$, the uncontrolled wait $X_k\to 0$ and  the external arrivals model becomes equivalent to a {\it generate-at-will} model in which a fresh arrival is admitted following the controlled waiting time $W_k=W'_k$.
In this limiting case, the sampling times $\{S_k\}$ are fully controlled.

Our goal is to design sampling policies, i.e., the controlled waiting times $\{W_k^\prime\}$, to optimize a weighted sum of the long-term average AoI and VoI. The long-term average AoI is defined as
\begin{align}
\overline{\texttt{AoI}}\triangleq \limsup_{N\rightarrow\infty}\frac{\sum_{k=1}^N\mathbb{E}\left[\int_{D_{k-1}}^{D_{k-1}+W_k+Y_k}\Delta(t)dt\right]}{\sum_{k=1}^N\mathbb{E}\left[D_k-D_{k-1}\right]},
\end{align}
which, from Figure~\ref{fig:samplepath}, can be more explicitly expressed as
\begin{IEEEeqnarray}{rCl} \label{eq_aoi}
\overline{\texttt{AoI}}=\limsup_{N\rightarrow\infty}\!\frac{
\sum_{k=1}^N\mathbb{E}\big[Y_{k-1}\left(W_k\!+\!Y_k\right)
\!+\!\tfrac{1}{2}
\left(W_k\!+\!Y_k\right)^2\big]}{\sum_{k=1}^N\mathbb{E}\left[D_k-D_{k-1}\right]}.
\IEEEeqnarraynumspace
\end{IEEEeqnarray}
The long-term average VoI on the other hand is defined as
\begin{align} \label{eq_voi}
\overline{\texttt{VoI}}\triangleq \limsup_{N\rightarrow\infty}\frac{\sum_{k=1}^N\mathbb{E}\left[\int_{D_{k-1}}^{D_{k-1}+W_k+Y_k}V(t)dt\right]}{\sum_{k=1}^N\mathbb{E}\left[D_k-D_{k-1}\right]}.
\end{align}

Next, we discuss the optimization problem in detail.

\section{Problem Formulation and Objective}

Our main goal is to choose the waiting times to minimize a convex combination of the long-term average AoI and long-term average negative VoI: 
\begin{align} \eqnlabel{opt_gen-main}
\min_{\{W'_k\geq0\}} \quad (1-\beta)\,\overline{\texttt{AoI}}
-\beta\,\overline{\texttt{VoI}},
\end{align}
for a
weighting factor $\beta$ satisfying $0\le\beta\le1$.

To get a handle on the above problem, let us denote by {\it epoch} $k$ the time elapsed in between the delivery of update $k-1$ and and the delivery of update $k$. During epoch $k$, value is being accumulated from update $k-1$, the most recently delivered update. The key question that we posed in the Introduction section now corresponds to how long should one accrue value from update $k-1$, given that it has class $i_{k-1}=i$ with value $\nu_{i}$ and value decay rate $\alpha_{i}$, before initiating the process to deliver update $k$.

Since processing times and class identities are i.i.d., we focus on {\it stationary deterministic} waiting policies, see, e.g., \cite{sample_quantize}. Specifically, when the most recent update has class $i_{k-1}=i$ and required processing time $Y_{k-1}$ the controlled waiting time in epoch $k$ is given by
\begin{subequations}
\label{eq_wait}
\begin{align} 
W'_k\triangleq w_i\left(Y_{k-1}\right),
\end{align}
where $w_i(\cdot)$ is a class-dependent waiting function to be optimized. That is, {\it the controlled waiting time in an epoch is a deterministic function of the previous epoch's processing time and sample value.} It follows from \eqnref{exogenousW} that the waiting time in epoch $k$ is 
\begin{align} \eqnlabel{exo:wait}
W_k\triangleq w_i\left(Y_{k-1}\right)
+X_k.\end{align}
\end{subequations}
Observe that such waiting policies induce a stationary distribution across epochs.

Substituting the waiting policy structure in \eqref{eq_wait} back in \eqref{eq_aoi} and \eqref{eq_voi}, the optimization problem in \eqnref{opt_gen-main} can be reduced to a functional optimization problem over a single epoch. Specifically, over epoch $k$ of duration $\TIME_k$, it can be seen from Figure~\ref{fig:samplepath} that the accumulated age and value are
\begin{align}
\AGE_k
&= Y_{k-1}\left(W_k+Y_k\right)+\frac{1}{2}\left(W_k+Y_k\right)^2,\eqnlabel{exo:AGEk}\\
\VALUE_k&= \int_{D_{k-1}}^{D_{k-1}+W_k+Y_k}\nu_{i_{k-1}}(t-S_{k-1})\,dt.
\eqnlabel{Vk-int1}
\end{align}
Since the expected duration of epoch $k$ is 
\begin{align}\eqnlabel{TIMEk}
\E{\TIME_k}=\E{W_k+Y_k},
\end{align}
the average  AoI and VoI become 
\begin{align}
 \AoI &=\frac{\E{\AGE_k}}{\E{\TIME_k}},& 
\VoI &=\frac{\E{\VALUE_k}}{\E{\TIME_k}}.
\eqnlabel{epochAoIVoI}
\end{align}

With the shorthand definition $\betabar\triangleq 1-\beta$, problem~\eqnref{opt_gen-main} now becomes
\begin{align} \label{opt_gen-epoch}
\min_{\{w_i(\cdot)\}}~~\frac{\betabar\Ebig{\AGE_k}-\beta\Ebig{\VALUE_k}}{\E{\TIME_k}}, \quad \mbox{s.t.} ~~ w_i(t)\geq0,~\forall t,i.
\end{align}

\section{Main Result}

In this section, we present our main result, namely the solution of problem \eqref{opt_gen-epoch}. Our solution will employ the Dinkelbach method \cite{dinkelbach-fractional-prog} which requires optimization over a free parameter $\theta$. Because the problem and solution are somewhat complicated, a summary of the optimal solution given in \Thmref{opt-wait} is presented in Section~\ref{sec_solution} and this is followed in Section~\ref{sec_discuss} by a discussion  of its basic properties. We then present some examples of the solution in Section~\ref{sec_examples}. The derivation of \Thmref{opt-wait} is deferred to Section~\ref{sec_derivation}.

\subsection{Solution Summary} \label{sec_solution}

Recalling that raw updates arrive at the processor as a rate $\lambda$ Poisson process, the uncontrolled wait $X_k$ is an exponential $(\lambda)$ random variable with moments $\E{X}=1/\lambda$, $\E{X^2}=2/\lambda^2$, and MGF $\Phi_X(s)=\lambda/(\lambda-s)$ . Since a class $i$ sample has processing time denoted $\Yhat_i$, the (overall) processing time $Y$ has moments
\begin{subequations}
\eqnlabel{Ymoments}
\begin{align}
\E{Y}&=\sum_i p_i\Ebig{\Yhat_i},\\ 
\E{Y^2}&=\sum_i p_i\Ebig{\Yhat_i^2}.
\end{align} 
\end{subequations}
It follows that the independent sum $Z=X+Y$ has moments
\begin{subequations}
\eqnlabel{Zmoments}
   \begin{align}
\E{Z}&= 1/\lambda + \E{Y},\eqnlabel{EZ}\\
\E{Z^2} &= 2/\lambda^2 +\frac{2}{\lambda}\E{Y}+\E{Y^2}.
\eqnlabel{EZ2}
\end{align} 
\end{subequations}
The Dinkelbach parameter $\theta$ specifies a threshold
\begin{align}
\thresh&\triangleq
    \theta-\betabar\E{Z}
\eqnlabel{summary:tau-threshold}
\end{align}
that is common to all classes.

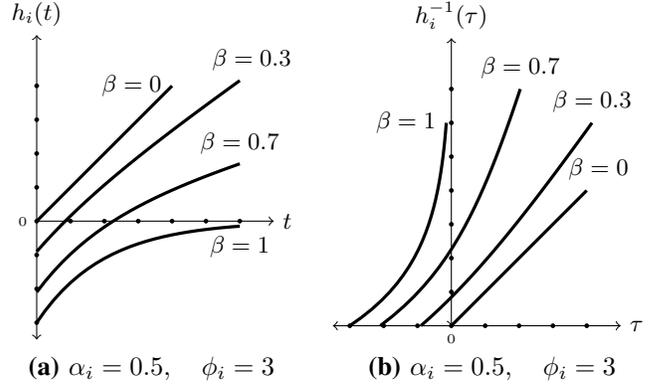
\begin{figure}[t]
\begin{tabular}{cc}
\begin{tikzpicture}[scale=0.45,baseline=(current bounding box.north)]
    \draw [<->,thin] (0,5.5) node [above] {\small $h_i(t)$} -- (0,-3.5);
\draw [->,thin] (0,0) -- (7,0) node [right] {\small$t$};
\fill
(0,0) circle[radius=2pt] node [left] {\tiny$0$}
(0,1) circle[radius=2pt] 
(0,2) circle[radius=2pt] 
(0,3) circle[radius=2pt] 
(0,4) circle[radius=2pt] 
(0,-1) circle[radius=2pt] 
(0,-2) circle[radius=2pt] 
(0,-3) circle[radius=2pt]  
(1,0) circle[radius=2pt] 
(2,0) circle[radius=2pt] 
(3,0) circle[radius=2pt] 
(4,0) circle[radius=2pt] 
(5,0) circle[radius=2pt] 
(6,0) circle[radius=2pt];
\draw[domain=0:4,very thick] plot (\x,\x) node [left] {\small$\beta=0$}; 
\draw[domain=0:6,very thick] plot (\x, {(0.7*\x)-(0.3*3*(exp(-0.5*\x)))})node [above] {\small\quad$\beta=0.3$}; 
\draw[domain=0:6,very thick] plot (\x, {(0.3*\x)-(0.7*3*(exp(-0.5*\x)))})node [above] {\small$\beta=0.7$};
\draw[domain=0:6,very thick] plot (\x, {-(3*exp(-0.5*\x)})node [below] {\small$\beta=1$};
\end{tikzpicture}
&
\begin{tikzpicture}[scale=0.45,baseline=(current bounding box.north)]
    \draw [<->] (0,8.5) node [above] {\small $h_i^{-1}(\tau)$} -- (0,0);
\draw [<->] (-3.5,0) -- (5,0) node [right] {\small$\tau$};
\fill
(0,0) circle[radius=2pt] node [below] {\tiny$0$}
(0,1) circle[radius=2pt] 
(0,2) circle[radius=2pt] 
(0,3) circle[radius=2pt] 
(0,4) circle[radius=2pt] 
(0,5) circle[radius=2pt] 
(0,6) circle[radius=2pt]
(0,7) circle[radius=2pt]
(-1,0) circle[radius=2pt] 
(-2,0) circle[radius=2pt] 
(-3,0) circle[radius=2pt] 
(1,0) circle[radius=2pt] 
(2,0) circle[radius=2pt] 
(3,0) circle[radius=2pt] 
(4,0) circle[radius=2pt];
\draw[domain=0:4,very thick] plot (\x,\x) node [above] {\small\quad$\beta=0$}; 
\draw[domain=0:6,very thick] plot ({(0.7*\x)-(0.3*3*(exp(-0.5*\x)))},\x)node [above] {\small$\beta=0.3$}; 
\draw[domain=0:7,very thick] plot ( {(0.3*\x)-(0.7*3*(exp(-0.5*\x)))},\x)node [above] {\small$\beta=0.7$};
\draw[domain=0:6,very thick] plot ( {-3*exp(-0.5*\x)},\x) node [left] {\small$\beta=1$};
\end{tikzpicture}\\
{\bf (a)} $\alpha_i=0.5,\quad \phi_i=3$
&{\bf (b)} $\alpha_i=0.5,\quad \phi_i=3$
\end{tabular}
\caption{Examples of (a) the threshold function $h_i(t)$ and (b) the corresponding inverses $h_i^{-1}(\tau)$. The minimum $h_i(0)=-\beta\phi_i$ occurs at $t=0$ and $h_i(t)\to(1-\beta)t$ as $t\to\infty$.}
\label{fig:hjinversefunctions}
\vspace{-.2in}
\end{figure}

The value delivered by a class $i$ update depends on its decay rate $\alpha_i$ through the {\em value decay factor} 
\begin{subequations}
\eqnlabel{opt-algorithm}
\begin{align}
\phi_i&\triangleq \nu_i\Phi_X(-\alpha_i)
\Phi_{\Yhat_i}(-\alpha_i)=\nu_i\frac{\lambda}{\lambda+\alpha_i}
\Phi_{\Yhat_i}(-\alpha_i),
\eqnlabel{phi-defn}
\end{align}
which is used to define the class $i$ threshold function
\begin{align}
h_i(t)\triangleq\betabar t-\beta\phi_i e^{-\alpha_i t},
\eqnlabel{hjdefn}
\end{align}
and this defines the class $i$ minimum inter-update time
\begin{align} \eqnlabel{summary:eq_yj}
    \ythresh\triangleq \begin{cases}
       h_i^{-1}(\thresh), & \tau(\theta)\ge -\beta\phi_i, \\
       0, & \ow.
    \end{cases}
\end{align}
Examples of these threshold functions and their inverses are shown in Figure~\ref{fig:hjinversefunctions}. 

When the {\em prior} processed update belongs to class $i$ and has service time $\Yhat_i$, the controlled waiting time during which arriving raw updates are discarded at the processor is 
\begin{align}
\eqnlabel{summary:opt-wait}
\What_i &=[\ythresh-\Yhat_i]^+.
\end{align}

For fixed $\theta$,
the expected accumulated age and value over one epoch are 
\begin{align}
  \Ebig{\AGE_k} &=
  \sum_i p_i\paren{(\ythresh+\E{Z})\Ebig{\What_i}-\Ebig{\What_i^2}/2}\nn 
  &\qquad + \E{Y}\E{Z}+\E{Z^2}/2,
\eqnlabel{summary:EAGEk}\\
\Ebig{\VALUE_{k}}
&=\sum_i\frac{p_i}{\alpha_i}\paren{\nu_i\Phi_{\Yhat_i}(-\alpha_i)-\phi_i\Phi_{\Yhat_i + \What_i}(-\alpha_{i})},\eqnlabel{summary:EVALUEk}
\end{align}
and the expected duration of an epoch is 
\begin{align}
\E{\TIME_k}&=\sum_ip_i\Ebig{\What_i}+\E{Z}.
\eqnlabel{summary:ETIMEk}
\end{align}
The average AoI and VoI are then given by \eqnref{epochAoIVoI}. 
With 
\begin{align} 
p(\theta)\triangleq \betabar\Ebig{\AGE_k}- \beta \Ebig{\VALUE_k}-\theta \E{\TIME_k},
\eqnlabel{opt_gen-theta} 
\end{align}
\end{subequations}
the following theorem describes how to find the optimal $\theta=\theta^*$ 
 that equals the minimum weighted AoI/VoI objective in \eqnref{opt_gen-main}. 
\begin{theorem}
\thmlabel{opt-wait}
The optimal waiting policy of problem~\eqnref{opt_gen-main} is given by the class-dependent threshold policy in \eqnref{summary:opt-wait}, with the class $i$ threshold $\ythresh$ in \eqnref{summary:eq_yj}. The value $\theta=\theta^*$ given by the unique solution of $p(\theta^*)=0$ in \eqnref{opt_gen-theta} is equal to the optimal solution of problem~\eqnref{opt_gen-main}.
\end{theorem}

In the derivation of the theorem in Section~\ref{sec_derivation}, we show that $\theta^*$ can be found by a simple bisection search.

\subsection{Discussion} \label{sec_discuss}

We observe that the solution of Theorem~\ref{thm:opt-wait} generalizes the AoI-optimal waiting strategy found in \cite{sun-age-mdp} for generate-at-will systems. In that work, when the prior service time is $y$, the source waits for time $w=[y^*-y]^+$ before generating the next update, for some threshold $y^*$.  Just like $y^*$ in 
\cite{sun-age-mdp}, $\ythresh$ can be interpreted as a minimum inter-update time. 

Because the threshold functions $h_i(t)$ are monotone increasing,  the minimum inter-update time $\ythresh$ increases  with $\theta$. Hence, increasing $\theta$ slows the processing rate of updates. Specifically, if the prior update was processed quickly and found to be in class $i$, the additional wait $\What_i$ is inserted before permitting an arriving update to be processed. This additional wait depends on the value decay factor $\phi_i$; {\it when the prior update is more valuable, the additional wait is larger.}

\section{Examples: Hyperexponential Service} \label{sec_examples}

Here we consider a multi-class system such that type $i$ samples occur with probability $p_i$, and have sample value  $\nu_i$ and decay rate $\alpha_i$. The service time is hyperexponential in that type $i$ samples have exponential $(\mu_i)$ processing times $\Yhat_i$. Hence,
\begin{align}
\Ebig{\Yhat_i}=\frac{1}{\mu_i},\quad
\Ebig{\Yhat_i^2}=\frac{2}{\mu_i^2},\quad \Phi_{\Yhat_i}(-\alpha_i)= \frac{\mu_i}{\mu_i+\alpha_i}.\eqnlabel{hyper:Yhat-moments}
\end{align}
From \eqnref{Ymoments}, it follows that the service time $Y$ has moments
\begin{align}
    \E{Y} &= \sum_i \frac{p_i}{\mu_i},\quad 
    \E{Y^2}= \sum_i \frac{2p_i}{\mu_i^2},
    \eqnlabel{hyper:Ymoments}
\end{align}
and $Z$ has moments given by \eqnref{Zmoments}.
With these values, \eqnref{phi-defn} yields the class $i$ value decay factor
\begin{subequations}
\begin{align}
\phi_i&=\nu_i\frac{\lambda}{\lambda+\alpha_i}\frac{\mu_i}{\mu_i+\alpha_i},
 \end{align}
 \end{subequations}
 while $h_i(t)$, $\thresh$, $\ythresh$,  and $\What_i$ are specified by \eqnref{hjdefn}-\eqnref{summary:opt-wait}. 
 Finding $\E{\AGE_k}$ in \eqnref{summary:EAGEk} requires the moments $\Ebig{\What_i}$ and $\Ebig{\What_i^2}$. With the shorthand notation $\ybar_i\triangleq \ythresh$,
 we use integration by parts to calculate
\begin{IEEEeqnarray}{rCl}
\Ebig{\What_i}&=&\int_0^{\ybar_i}
    \!(\ybar_i-y)\mu_ie^{-\mu_i y}\,dy
    = \ybar_i-\frac{1}{\mu_i}+\frac{e^{-\mu_i\ybar_i}}{\mu_i},\IEEEeqnarraynumspace
    \eqnlabel{hyper:EW}\\
\Ebig{\What_i^2}&=&\int_0^{\ybar_i} (\ybar_i-y)^2\mu_ie^{-\mu_i y}\,dy
=\ybar_i^2-\frac{2\Ebig{\What_i}}{\mu_i}.
\eqnlabel{hyper:EW2}
\end{IEEEeqnarray} 
Applying \eqnref{hyper:EW2} to \eqnref{summary:EAGEk} yields
\begin{align}
  \Ebig{\AGE_k} &=
  \sum_i p_i\paren{(\ybar_i+1/\mu_i+\E{Z})\Ebig{\What_i}-\ybar_i^2/2}\nn 
  &\qquad + \E{Y}\E{Z}+\E{Z^2}/2.
\eqnlabel{hyper:EAGEk}
\end{align}
Substituting $\E{Y}$ from \eqnref{hyper:Ymoments} and $\Ebig{\What_i}$ from \eqnref{hyper:EW}
in \eqnref{hyper:EAGEk} yields
\begin{align}
  \Ebig{\AGE_k} &=
  \sum_i p_i\bigg(\ybar_i^2/2-\frac{1}{\mu_i^2} +\ybar_i\E{Z}\nn
  &\qquad+(\ybar_i+\frac{1}{\mu_i}+\E{Z})\frac{e^{-\mu_i\ybar_i}}{\mu_i}\bigg)+ \frac{\Ebig{Z^2}}{2}.
\eqnlabel{hyper:EAGEk-v2}
\end{align}  
From \eqnref{EZ2} and \eqnref{hyper:Ymoments}, we obtain
\begin{align}
    \E{Z^2}=\frac{2}{\lambda^2} +2\sum_i p_i\paren{\frac{1}{\lambda\mu_i}+\frac{1}{\mu_i^2}}.
\end{align}

\begin{figure}[t]
\centering\includegraphics{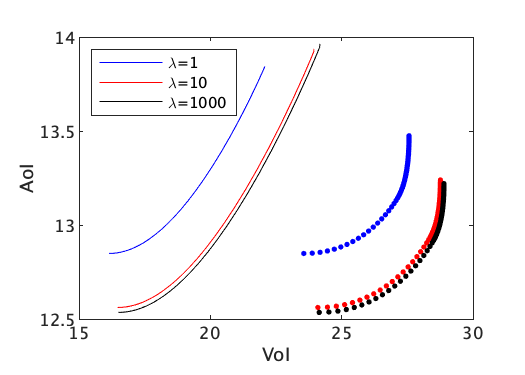}
\caption{AoI vs.~VoI tradeoff for a two class system with hyperexponential service: $[p_1,p_2]=[0.5,0.5]$, $[\nu_1, \nu_2]= [100,1]$, $[\alpha_1, \alpha_2]= [0.1,1]$ and  $[\mu_1,\mu_2]=[0.1,1]$ (solid lines) or $[\mu_1,\mu_2]=[1,0.1]$ (dotted lines).}
\label{fig:hyper:AoIVoI}
\vspace{-.1in}
\end{figure}
It follows that
\begin{align}
  \Ebig{\AGE_k} &=
  \sum_i p_i\bigg(\ybar_i^2/2+\frac{1}{\lambda\mu_i} +\ybar_i\E{Z}\nn
&\qquad+\big(\ybar_i+\frac{1}{\mu_i}+\E{Z}\big)\frac{e^{-\mu_i\ybar_i}}{\mu_i}\bigg)+\frac{1}{\lambda^2}.
\eqnlabel{hyper:EAGEk-v3}
\end{align}  

To find $\E{\VALUE_k}$, we need 
$\Phi_{\Yhat_i+\What_i}(-\alpha_i)$.
From the observation $
\Yhat_i+\What_i=\max(\ybar_i,\Yhat_i)$,
we can write
\begin{align}
    \Phi_{\Yhat_i+\What_i}(s)
    &=\E{e^{s\max(\ybar_i,\Yhat_i)}}\nn
    &=e^{s\ybar_i}\prob{\Yhat_i\le \ybar_i}+\int_{\ybar_i}^\infty e^{sy}\mu_ie^{-\mu_i y}\,dy\nn
    &=e^{s\ybar_i}(1-e^{\mu_i\ybar_i})
+\frac{\mu_i}{\mu_i-s}e^{-(\mu_i-s)\ybar_i}.\eqnlabel{hyper:YW-MGF}
\end{align}
Substituting 
 \eqnref{hyper:Yhat-moments} and  \eqnref{hyper:YW-MGF} with $s=-\alpha_i$ into \eqnref{summary:EVALUEk} yields
\begin{IEEEeqnarray}{rcl}
    \Ebig{\VALUE_{k}}
&=&\sum_i\frac{p_i\nu_i\mu_i}{\alpha_i(\mu_i+\alpha_i)}
\bracket{1\!-\!\frac{\lambda e^{-\alpha_i\ybar_i}}{\lambda+\alpha_i}\bigg(1\!-\!\frac{\alpha_i e^{-\mu_i\ybar_i}}{\mu_i+\alpha_i}\bigg)}.
\eqnlabel{hyper:EVALUEk}
\IEEEeqnarraynumspace
\end{IEEEeqnarray}
Furthermore, we see from \eqnref{summary:ETIMEk} and \eqnref{hyper:EW}
that
\begin{align}
    \E{\TIME_k}&=\sum_i p_i\bigg[\ybar_i+\frac{e^{-\mu_i\ybar_i}}{\mu_i}\bigg] +\frac{1}{\lambda}.\eqnlabel{hyper:ETIMEk}
\end{align} 
    
The selection of $\theta$ specifies the threshold $\thresh$ and the minimum wait $\ybar_i=\ythresh$. The AoI and VoI and the weighted combination $p(\theta)$ in \eqnref{opt_gen-theta} can then be calculated from $\E{\AGE_k}$ in \eqnref{hyper:EAGEk}, $\E{\VALUE_k}$ in \eqnref{hyper:EVALUEk}, and $\E{\TIME_k}$ in \eqnref{hyper:ETIMEk}. Thus, for each value of $\beta\in[0,1)$, we search for $\theta^*$ satisfying $p(\theta^*)=0$ in order to find the minimum combined AoI and VoI in \eqnref{epochAoIVoI}. That is, at $\theta=\theta^*$, we obtain an optimal AoI and VoI pair. By varying $\beta$ and finding the $\theta^*$ for each $\beta$, we obtain the boundary of feasible (AoI,VoI) pairs. 

In the example of Figure~\ref{fig:hyper:AoIVoI}, class $1$ and class $2$ samples  arrive equiprobably but with $[\nu_1,\nu_2]=[100,1]$;  class $1$ samples are $100\times$ more valuable. Moreover, with $[\alpha_1,\alpha_2]=[0.1,1]$, the class 1 value decays $10\times$ more slowly. The figure depicts two collections  (solid/dotted lines) of AoI/VoI tradeoff curves. In the solid line case, $[\mu_1,\mu_2]=[0.1,1]$, corresponding to the valuable updates requiring $10\times$ greater processing time. In the dotted line case, $[\mu_1,\mu_2]=[1,0.1]$, so the valuable samples are now the updates that are processed quickly.  
 In both cases, we see that we can choose whether to minimize age or to maximize value by varying the weight $\beta$. We also see that increasing the sample arrival rate $\lambda$ improves the tradeoff by enabling more precise control of when to start processing the next sample. In both cases, $\lambda=10$ and $\lambda=1000$ are almost the same and $\lambda=1000$ is essentially indistinguishable from  the generate-at-will system performance. We note, however, that the dotted line curves offer better AoI/VoI tradeoffs simply because processing the valuable updates fast results in less value being lost in  processing.

\begin{figure}[t]
\centering\includegraphics{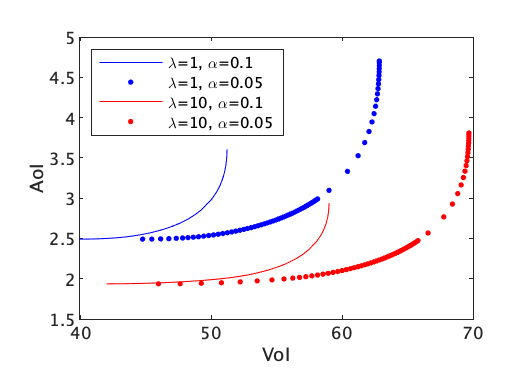}
\caption{AoI vs.~VoI tradeoff for a two class system with hyperexponential service: $[p_1,p_2]=[0.5,0.5]$, $[\nu_1, \nu_2]= [100,1]$, $[\mu_1,\mu_2]=[1,1]$ and $[\alpha_1, \alpha_2]= [0.1,0.1]$ (solid lines) or $[\alpha_1,\alpha_2]=[0.05,0.05]$ (dotted lines).}
\label{fig:hyper:AoIVoI-alpha}
\vspace{-.2in}
\end{figure}

A second example in Figure~\ref{fig:hyper:AoIVoI-alpha} examines the effect of decay parameters $\alpha_i$. Specifically, the two classes have the same decay rate $\alpha_1=\alpha_2=\alpha$ and the figure compares $\alpha=0.1$ against $\alpha=0.05$. As in the previous example, class 1 and 2 samples are equiprobable and class 1 samples have  $100\times$ higher value.  As we would expect, reducing the decay rate $\alpha$ increases the VoI at the monitor and improves the AoI/VoI tradeoff. Similarly, increasing the sample arrival rate $\lambda$ improves the AoI-VoI tradeoff.

\section{Derivation of Theorem~\ref{thm:opt-wait}} \label{sec_derivation}

Over the interval $[D_{k-1},D_k]$, value of information is accrued from update $k-1$, which was sampled at time $S_{k-1}$ and delivered to the monitor at time \begin{equation}
D_{k-1}=S_{k-1}+Y_{k-1}. \end{equation} 
With exogenous arrivals and decaying exponential value functions, it follows from \eqnref{Vk-int1} and the substitution $\tau=t-D_{k-1}$ that 
the accumulated VoI over the interval is
\begin{IEEEeqnarray}{rCl}
\VALUE_k   
&=&\int_0^{W_k+Y_k}\nu_{i_{k-1}}e^{-\alpha_{i_{k-1}}(\tau+Y_{k-1})}\, d\tau\nn
&=&\frac{\nu_{i_{k-1}}}{\alpha_{i_{k-1}}}e^{-\alpha_{i_{k-1}}Y_{k-1}}\paren{1-e^{-\alpha_{i_{k-1}}(W_k+Y_k)}}.
\eqnlabel{VALUEk2}
\end{IEEEeqnarray}
When update $k-1$ has class $i_{k-1}=i$, we denote the update processing time as $Y_{k-1}=\Yhat_i$ to highlight its known class dependence. Similarly,  we denote the class-dependent controlled wait by $\What_i=w_i(\Yhat_i)$. With this notation, \eqnref{exo:wait} becomes
\begin{align}\eqnlabel{exo:Wk}
  W_k= \What_i + X_k. 
\end{align}
Averaging over the class $i_{k-1}$ of update $k-1$,   \eqnref{VALUEk2} yields the expected value
\begin{IEEEeqnarray}{rCl}
\Ebig{\VALUE_{k}}
&=&\sum_i\frac{p_i\nu_i}{\alpha_i}\E{e^{-\alpha_i
\Yhat_i}\paren{1-e^{-\alpha_{i}(\What_i+X_k+Y_k)}}}.
\IEEEeqnarraynumspace
\end{IEEEeqnarray}
With the value decay factor $\phi_i$ defined in \eqnref{phi-defn}, it follows from mutual independence of $(\hat{W}_i,\hat{Y}_i)$, $X_k$ and $Y_k$ that $\Esmall{\VALUE_k}$ is given by \eqnref{summary:EVALUEk}.

Over the same epoch, \eqnref{exo:AGEk}, \eqnref{exo:Wk} and the definition
$Z_k\triangleq X_k+Y_k$ imply that the expected accumulated age is
\begin{align}
   \Ebig{\AGE_k} &=\sum_i p_i\Ebig{\Yhat_{i}(\What_i+Z_k) 
   +\frac{1}{2}\big(\What_i+Z_k\big)^2}.\eqnlabel{EAGEk}
   \end{align}
Since $Z_k$ is independent of $\Yhat_i$ and $\What_i$, we have
\begin{align}
    \Ebig{\AGE_k} &=
  \sum_i p_i\Ebigg{\Yhat_i\What_i +\What_i^2/{2} +\What_i\E{Z}}\nn 
  &\qquad + \E{Y}\E{Z}+\E{Z^2}/2.
\eqnlabel{proof:EAGEk}
\end{align}
In addition, we observe from \eqnref{TIMEk}, \eqnref{exo:Wk}, and the definition of $Z_k$ that $\E{\TIME_k}$ is given by \eqnref{summary:ETIMEk}.

We now follow Dinkelbach's approach \cite{dinkelbach-fractional-prog} to transform problem \eqref{opt_gen-epoch} into the following auxiliary problem:
\begin{subequations}\eqnlabel{exo:opt_gen-aux}
\begin{align} 
p(\theta)\triangleq \min_{\set{w_i(\cdot)}}& \quad \betabar\Ebig{\AGE_k}- \beta \Ebig{\VALUE_k}-\theta \E{\TIME_k}\\
\mbox{s.t.~}& \quad w_i(t)\geq 0,~\forall i,t,
\end{align}
\end{subequations}
for some $\theta\in\mathbb{R}$. The solution of problem \eqref{opt_gen-epoch} is now given by the unique $\theta^*$ that solves $p(\theta^*)=0$ \cite{dinkelbach-fractional-prog}, which can be found by, e.g., a bisection search.

To solve problem~\eqnref{exo:opt_gen-aux}, we define the Lagrangian
\begin{IEEEeqnarray}{rCl}
\mathcal{L}&=&\betabar\Ebig{\AGE_k}-\beta\Ebig{\VALUE_k}-\theta\E{\TIME_k}-\!\sum_i\!\!\int\!\! w_i(y)\eta_i(y)\,dy,
    \IEEEeqnarraynumspace
\end{IEEEeqnarray}
where $\eta_i(y)$ is a Lagrange multiplier. It follows from  \eqnref{proof:EAGEk}, \eqnref{summary:EVALUEk},  and \eqnref{summary:ETIMEk} that
\begin{IEEEeqnarray}{rCl}
\mathcal{L}&=&\sum_ip_i\int\!\!\fyi(y)\bigg[ \betabar yw_i(y)+ \frac{\betabar w^2_i(y)}{2} +w_i(y)(\betabar\E{Z}-\theta)\nn
&&\hspace{1.8in} +\beta\frac{\phi_i}{\alpha_i}e^{-\alpha_i[y+w_i(y)]}\bigg]\,dy 
\nonumber \\
&&-\sum_i\int w_i(y)\eta_i(y)\,dy+\betabar\E{Y}\E{Z}+\tfrac{1}{2}\betabar\E{Z^2}\nn
&&-\beta\sum_i\frac{p_i}{\alpha_i}\nu_i\Phi_{\Yhat_i}(-\alpha_i) -\theta\E{Z},
\end{IEEEeqnarray}
where $\hat{f}_i(y)$ denotes the density of $\hat{Y}_i$. Taking the functional derivative of $\mathcal{L}$ with respect to $w_j(y)$ and equating to $0$ yields
\begin{align}
    &\betabar[y+ w_j(y)]  -\beta\phi_je^{-\alpha_j(y+w_j(y))}\nn   &\hspace{1.5in}=\frac{\eta_j(y)}{p_j\fyi[j](y)}+\theta-\betabar\E{Z}.
\eqnlabel{Lderiv0}
\end{align}
From the definition of the threshold function
$h_j(t)$ in \eqnref{hjdefn} and the threshold $\thresh$ in \eqnref{summary:tau-threshold}, the condition~\eqnref{Lderiv0} becomes
\begin{IEEEeqnarray}{rCl}
h_j(y+w_j(y))&=&\frac{\eta_j(y)}{p_j\fyi[j](y)}+\thresh.
\IEEEeqnarraynumspace\eqnlabel{Lderiv-h}
\end{IEEEeqnarray}
We observe that $h_j(t)$ is a strictly increasing function  and thus has an inverse $h_j^{-1}(\cdot)$.
Moreover, for $w\ge 0$, 
\begin{align} \label{eq_hj}
h_j(y+w)\ge h_j(y)&= \betabar y-\beta\phi_je^{-\alpha_j y}.
\end{align}
Hence, if $h_j(y)>\thresh$, or equivalently
\begin{align} \label{eq_yj}
    y> \ythresh[j]\triangleq h_j^{-1}(\thresh),
\end{align}
then  \eqnref{Lderiv-h} is satisfied with $w_j(y)=0$ and $\eta_j(y)>0$. Otherwise, \eqnref{Lderiv-h} is satisfied with $\eta_j(y)=0$ and $w_j(y)$ such that 
$y+w_j(y)=\ythresh[j]$. Putting these facts together, the optimal controlled-wait function is $w^*_j(y)=[\ythresh[j]-y]^+$. Returning to \eqnref{proof:EAGEk}, we have shown that $\What_i=[\ythresh-\Yhat_i]^+$ and this implies
\begin{align}
    \E{\Yhat_i\What_i}
    &=\E{(\ythresh-[\ythresh-\Yhat_i])[\ythresh-\Yhat_i]^+}\nn
    &=\E{\ythresh \What_i-\What_i^2}.
    \eqnlabel{proof:EYW}
\end{align}
Substituting \eqnref{proof:EYW} in \eqnref{proof:EAGEk} yields \eqnref{summary:EAGEk}.
This concludes the derivation of Theorem~\ref{thm:opt-wait}.

\bibliographystyle{unsrt}
\bibliography{AOI-2020-05copy}

\end{document}